\documentclass[aps,preprint,showpacs,epsfig] {revtex4}
\usepackage{pdfsync}
\usepackage{graphicx}   
\usepackage{epstopdf}
\usepackage{amsmath}

\begin{document}

\title{The Surface Electroclinic Effect near the First Order Smectic-$A^*$--Smectic-$C^*$ transition}

\author{Karl Saunders $^1$$^*$ and Per Rudquist $^2$  }

\affiliation{$^1$ Department of Physics, California Polytechnic State
University, San Luis Obispo, CA 93407, USA}

\affiliation{$^2$ Microtechnology and Nanoscience, Chalmers University of Technology, 412 96 G\"{o}teborg
Sweden}

\email{ksaunder@calpoly.edu}

\date{\today}


\begin{abstract}

We analyze the surface electroclinic effect (SECE) in a material that exhibits a first order bulk smectic-$A^*$ (Sm-$A^*$) -- smectic-$C^*$ (Sm-$C^*$) transition. The effect of a continuously varying degree of enantiomeric excess on the SECE is also investigated. We show that due to the first order nature of the bulk Sm-$A^*$ -- Sm-$C^*$ transition, the SECE can be unusually strong and that as enantiomeric excess is varied, a jump in surface induced tilt is expected. A theoretical state map, in enantiomeric excess - temperature space, features a critical point which terminates a line of first order discontinuities in the surface induced tilt. This critical point is analogous to that found for the phase diagram (in electric field - temperature space) for the bulk electroclinic effect. Analysis of the decay of the surface induced tilt, as one moves from surface into bulk shows that for sufficiently high surface tilt the decay will exhibit a well defined spatial kink within which it becomes especially rapid. 
We also propose that the SECE is additionally enhanced by the de Vries nature (i.e. small layer shrinkage at the bulk Sm-$A^*$ -- Sm-$C^*$ transition) of the material. As such the SECE provides a new means to characterize the de Vries nature of a material. We discuss the implications for using these materials in device applications and propose ways to investigate the predicted features experimentally.

\end{abstract}


\maketitle

\section{Introduction}

\begin{figure}
\begin{center}
\includegraphics[scale=0.4]{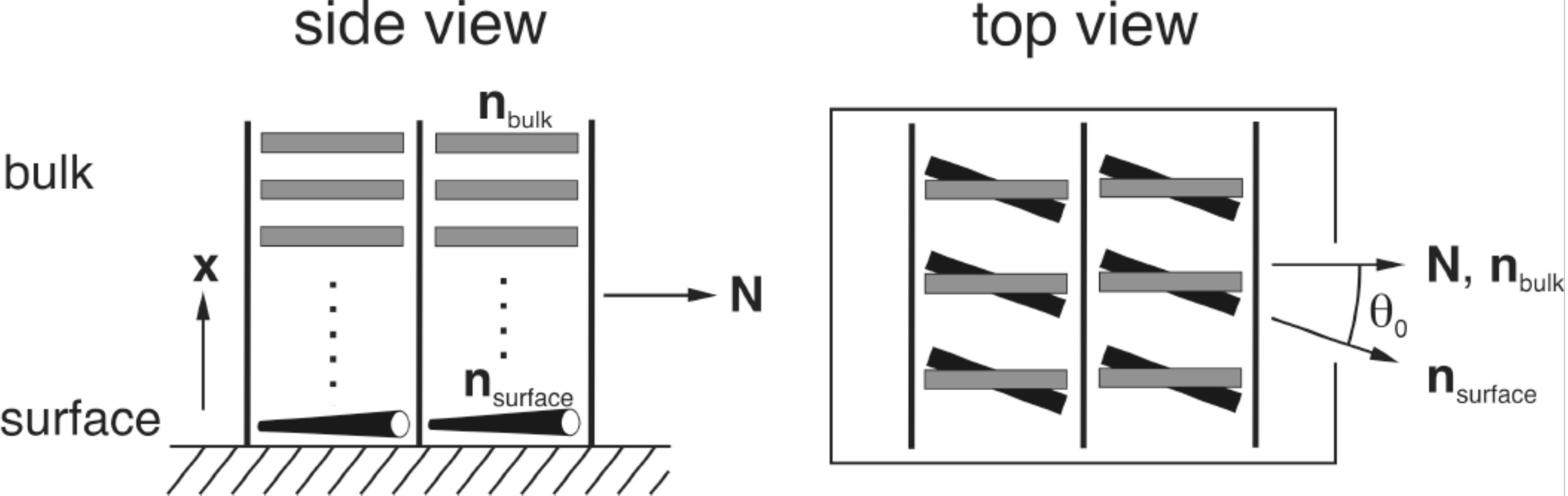}
\caption{A schematic of the surface electroclinic effect (SECE), whereby a coupling between molecular dipoles and a surface induces local tilt of the director ${\bf \hat n}$, away from the smectic layer normal ${\bf \hat N}$ at the surface. Conversely, on rubbed surfaces the SECE makes the bulk smectic normal ${\bf \hat N}$ form at an angle to the rubbing direction ${\bf \hat n_{surface}}$.}
\label{SECESchematic}
\end{center}
\end{figure}

The surface electroclinic effect (SECE) is an experimentally observed phenomenon in the chiral smectic-A (Sm-$A^*$) phase whereby a coupling between molecular dipoles and a surface induces local tilt of the director ${\bf \hat n}$, away from the smectic layer normal ${\bf \hat N}$ at the surface, as shown in Fig.~\ref{SECESchematic}. The SECE has been analyzed, both experimentally and theoretically \cite{Xue&Clark,Chen&Ouchi,Rovsek&Zeks, Shao&Boulder}, for materials with a continuous Sm-$A^*$ -- Sm-$C^*$ transition. The investigation of W415 \cite{Shao&Boulder} was notable in that an unusually large SECE was found and the effects of varying the enantiomeric excess were explored. The SECE has not however, been analyzed for materials with first order Sm-$A^*$ -- Sm-$C^*$ transitions \cite{415Footnote}. One motivation for the investigation of the SECE in materials with first order Sm-$A^*$ -- Sm-$C^*$ transitions is that the strong bulk electroclinic effect (BECE) of some de Vries materials can be attributed to the first order nature of the Sm-$A^*$ -- Sm-$C^*$ transition \cite{Saunders,EC Footnote}. The strong BECE with its fast analog electrooptic characteristics makes such materials technologically promising. However, as will be discussed below, the first order nature of the transition may also lead to a strong SECE, which has implications for the design of proper surface alignment layers and possibly also for the long-term stability of the alignment in, for instance, ferroelectric liquid crystal display devices based on de Vries materials.

Experimental and theoretical work \cite{Bahr&Heppke1,Bahr&Heppke2} on the BECE in materials with first order Sm-$A^*$ -- Sm-$C^*$ transitions demonstrated the existence of a critical point (in field ($E$) -- temperature ($T$) parameter space) which terminates a line of first order Sm-$C^*$ -- Sm-$C^*$ transitions. For the SECE the analogous parameters would be surface coupling and temperature, although it is not immediately obvious how one can explore such a parameter space experimentally. In sample cells in which one of the two polymer coated glass plates is rubbed to align the director ${\bf \hat n}$ at the surface (as shown in Fig. \ref{SECESchematic}) the SECE makes the smectic layer normal  ${\bf \hat N}$ deviate by an angle $\theta_0$ from the rubbing direction, i.e. from ${\bf \hat n_{surface}}$. This angle $\theta_0$ can be measured using polarized light microscopy.  However, surface pinning prevents the once-formed layer structure from rotating, which means that the surface tilt angle $\theta_0$ is effectively stuck at the temperature, $T_A$, where the layers first form in the Sm-$A^*$ phase \cite{Chen&Ouchi}. Thus, for such sample cells one cannot explore the variation of the SECE with $T$. We note that $T_A$ could perhaps be varied by quenching the system into the Sm-$A^*$ phase at lower temperatures. Another approach would be to establish an alignment direction for ${\bf \hat N}$ without doing so for ${\bf \hat n}$, i.e. align the smectic layers in a homogeneous bookshelf structure without having a rubbing direction at the surfaces. This would then allow ${\bf \hat n}$ to rotate, and thus $\theta_0$ to change, at the surface as temperature is varied. One way to do this is to shear-align the sample, i.e. to slide the top substrate with respect to the lower substrate under an applied ac field, as was done with the first experiments on surface-stabilized ferroelectric liquid crystal cells \cite{Shear Alignment}.

The other option is to vary the surface coupling, which increases monotonically with enantiomeric excess $\epsilon$. This should be done by preparing separate cells with varying degrees of $\epsilon$. It could in principle also be achieved via a cell made up of a contact preparation between opposite-handed moleclules of the same compound. Keeping the cell at elevated temperatures (i.e., in the isotropic phase) after filling will facilitate diffusion over the contact boundary until a gradient in $\epsilon$ develops over a suitable distance in the cell plane. On one side of such a cell $\epsilon>0$ and as one moves to the center of the cell (where there is an equal number of left and right handed molecules, resulting in a racemic mixture) $\epsilon\rightarrow 0$. The second half of the cell is the mirror image of the first half, with $\epsilon<0$. After a long enough time continued diffusion will make the gradient in $\epsilon$ gradually disappear,  but immediately after the first cooling down from the isotropic phase the surface coupling varies continuously across the cell in the ${\bf \hat y}$ direction, where ${\bf \hat y}$ is perpendicular to both ${\bf \hat N}$ and ${\bf \hat x}$. Provided the gradient in $\epsilon (y)$ is known we could directly probe the $\theta_0(\epsilon (y))$ for all values of $\epsilon$. In such an experiment, care should be taken to avoid any influence on the structure from the anchoring on the opposing surface. Moreover, with time the gradient in $\epsilon(y)$ on the surface might differ from the one in the bulk, due to surface pinning effects.

As part of the present analysis we obtain a theoretical state map, as shown in Fig.~\ref{StateMap}(a), in enantiomeric excess - temperature space which features a critical point that terminates a line of first order discontinuities in the surface induced tilt. This critical point ($\epsilon_c$, $T_c$) is analogous to that found for the phase diagram (in electric field - $T$ space) for the BECE. We will demonstrate that for sufficiently low temperature $T<T_c$, there will be a jump in surface induced tilt as $\epsilon$ is varied, as shown in Fig.~\ref{StateMap}(b). This jump is analogous to the field-induced first order Sm-$C^*$ -- Sm-$C^*$ transition for the BECE. As with the BECE, the SECE is enhanced due to the first order nature of the zero-field Sm-$A^*$ -- Sm-$C^*$ transition \cite{EC Footnote}. However, we will also argue that the SECE in de Vries type materials is further enhanced due to the minimal layer spacing change at the Sm-$A^*$ -- Sm-$C^*$ transition. Recent work \cite{Saunders, Lemieux}, indicates that first order Sm-$A^*$ -- Sm-$C^*$ transitions with minimal layer chance are hallmarks of de Vries-type materials. Thus, an enhanced SECE effect in a material (such as that described here) could be a signature of its de Vries-like nature.

We also investigate how the surface induced tilt decays as one moves from the surface into the bulk. For materials with continuous Sm-$A^*$ -- Sm-$C^*$ transitions, such a decay is gentle, meaning that the slope of the decay decreases the further one moves into the bulk. Our analysis shows that for a system with a first order Sm-$A^*$ -- Sm-$C^*$ transition this will not be the case and that there will be a kink in the decay. The kink would be a region in which the the slope becomes steeper and the decay becomes much more rapid. This is shown in Fig.~\ref{StateMap}(c).

\section{Model and Analysis}

In this article we consider a cell with fixed, uniform enantiomeric excess $\epsilon>0$. We will present analysis for a spatially varying $\epsilon$ (as in the contact preparation discussed above) in a future publication \cite{Rudquist&Saunders}. We consider a standard, simplified, geometry (shown in Fig.~\ref{SECESchematic}) with the liquid crystal in contact with a single surface at $x=0$ and extending to $x=\infty$. This is valid if the two cell surfaces are separated by a distance $L_x\gg\xi_x$, where $\xi_x$ is a correlation length to be defined below. With this geometry the layer normal ${\bf \hat N}$ will tilt away from the rubbing direction ${\bf \hat n_{surface}}$, within the $yz$ plane, by an amount $\theta(x)$. This tilt is biggest at the surface, i.e. at $x=0$, and decays to zero with increasing $x$, i.e. $\theta(x\rightarrow\infty)=0$. To analyze the SECE in the Sm-$A^*$ phase we employ a Landau expansion of the bulk and surface free energies, $F_B$ and $F_S$ respectively:
\begin{eqnarray}
F_B= A_\perp \int_0^{\infty} dx \left[f(\theta) + \frac{1}{2\chi} P^2 - \gamma\epsilon P \theta \right]\;
\label{Bulk free energy}
\end{eqnarray}
and
\begin{eqnarray}
F_S= -A_\perp  s P(x=0)\;,
\label{Surface free energy}
\end{eqnarray}
where $A_\perp$ is the area of the surface, $P(x)$ is the component of the average polarization perpendicular to the surface and $\gamma$ is a $\theta$-$P$ coupling constant. The strength of the $\theta$-$P$ coupling is proportional to $\epsilon$. The coefficient $\chi$ is a generalized d.c. electric susceptibility and the coefficient $s$ is proportional to the strength of the polar surface anchoring. In keeping only terms of order $P^2$ we make the standard assumption that the Sm-$A^*$ - Sm-$C^*$ transition is primarily driven by $\theta$, with $P$ playing a secondary role. 

The piece $f(\theta)$ is given by:  
\begin{eqnarray}
f(\theta)=\frac{1}{2} a(T) \theta^2 + \frac{1}{4} b \theta^4+\frac{1}{6} c \theta^6 + \frac{1}{2} K_x \left({\partial_x\theta}\right)^2\;,
\label{f_u}
\end{eqnarray}
where $a(T)=r(T-T_0)/T_0$ and $c>0$. For $b\geq0$ the racemic, i.e. $\epsilon=0$, bulk Sm-$A$ - Sm-$C$ transition is continuous and takes place at $T_{2nd}=T_0$ while for $b<0$ the transition is first order and takes place at $T_{1st}=T_0\left( 1+\frac{3 b^2}{16 c r}\right)$. We consider a racemic bulk Sm-$A$ - Sm-$C$ transition that is first order, i.e., $b<0$. It should be pointed out that the transition can also be driven first order by increasing the enantiomeric excess \cite{Liu&Huang&Min}. For the sake of simplicity, we do not consider that possibility here but will do so in a future publication \cite{Rudquist&Saunders}. To lowest order $K_x=K_T$, the twist elastic constant, and controls the decay, over a length scale $\xi_x=\sqrt{K_T/a}$, of $\theta$ along $x$ into the bulk. We do not include elastic energy contributions due to the spatial variation (along $x$) of the layer spacing. For de Vries-type materials, the tilt induced layer contraction will be minimal. 

Setting $P$ equal to its minimum value $\chi \gamma \epsilon  \theta$ leads to an elimination of the last two terms of Eq.~(\ref{Bulk free energy}) and an $\epsilon$ dependent  $a(T,\epsilon)=r\left(\frac{T}{T_0}-1-\frac{\chi\gamma^2\epsilon^2}{r}\right)$ and $\xi_x(T,\epsilon)=\sqrt{K_T/a(T,\epsilon)}$. There is a corresponding 
upward, $\epsilon$ dependent, renormalization of $T_{1st}^*(\epsilon)$:
\begin{eqnarray}
T_{1st}^*(\epsilon)= T_{1st}\left( 1+\frac{\chi\gamma^2\epsilon^2}{r}\frac{T_0}{T_{1st}}\right)\;.
\label{T_1st}
\end{eqnarray}
The dependence of $T_{1st}^*(\epsilon)$ on $\epsilon$ means that if $\epsilon$ is sufficiently large then the system's bulk will actually be in the Sm-$C^*$ phase. Alternatively, for the case of fixed temperature $T$ it is useful to express the value of $\epsilon$ above which the system's bulk will be in the Sm-$C^*$ phase:
\begin{eqnarray}
\epsilon_{1st}(T)=\sqrt{r(T-T_{1st})/\chi T_0}/\gamma\;.
\label{epsilon_1st}
\end{eqnarray}

\begin{figure}
\begin{center}
\includegraphics[scale=0.7]{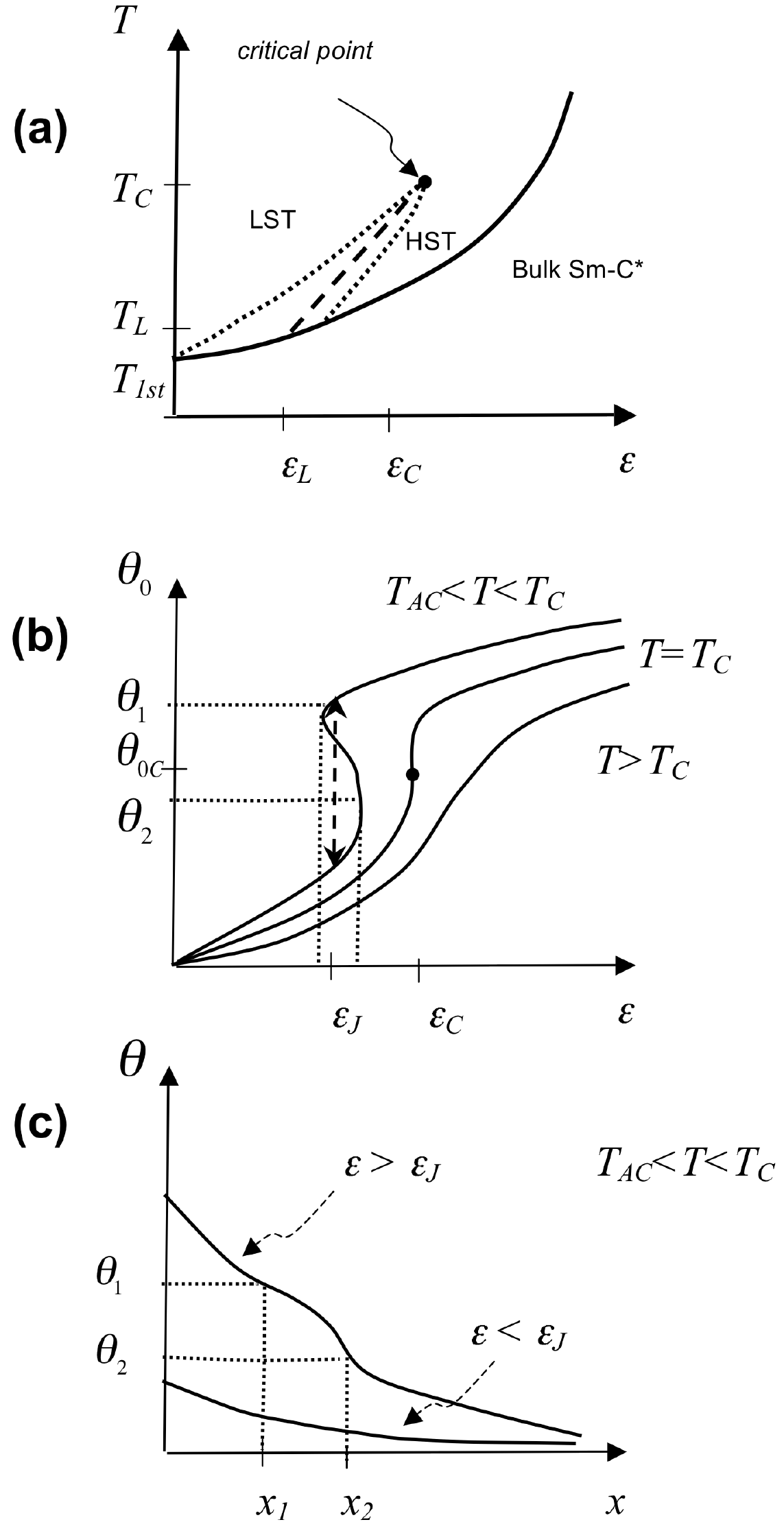}
\caption{(a) Surface electroclinic effect state map in enantiomeric excess ($\epsilon$) -- temperature ($T$) parameter space. See text for a detailed explanation of the state map. (b) Three fixed-temperature profiles of $\theta_0(\epsilon)$. Only $\epsilon$ values corresponding to the system's bulk being in the Sm-$A^*$ phase are considered. See text for a detailed explanation of these profiles. (c) Decay of tilt from the surface into the bulk, i.e. along the $x$ direction for $T<T_C$. Only values $\epsilon<\epsilon_{1st}$ are considered, i.e. the bulk of the system is in the Sm-$A^*$ phase, where the tilt is zero. For low surface tilt (LST) the decay of $\theta$ does not have a kink. For high surface tilt (HST) the decay of the tilt does have a kink in an intermediate range of $x$. }
\label{StateMap}
\end{center}
\end{figure}

It is useful to rescale both $\theta \rightarrow  \sqrt{\frac{|b|}{2c}} \theta$ and $x \rightarrow \xi_x x$. Doing so leads to the following rescaled bulk and surface free energies:
\begin{eqnarray}
F_B= A_\perp\frac{|b|^3}{16c^2}f_B= A_\perp\frac{|b|^3}{16c^2} \xi_x \int_0^{\infty} dx\left[ \alpha(T,\epsilon) \theta^2 -  \theta^4+\frac{1}{3} \theta^6+\alpha(T,\epsilon)\left({\partial_x\theta}\right)^2 \right]\;
\label{Rescaled free energy}
\end{eqnarray}
and 
\begin{eqnarray}
F_S= A_\perp\frac{|b|^3}{16c^2}f_S= -A_\perp\frac{|b|^3}{16c^2} \left[8 s \chi\gamma\epsilon \sqrt{\frac{2c^3}{|b|^5}} \theta_0\right]\;,
\label{Rescaled free energy}
\end{eqnarray}
where $f_B$ and $f_S$ are the rescaled bulk and surface free energies per unit area, $\alpha(T,\epsilon)=\frac{4a(T,\epsilon)c}{|b|^2}$, and $\theta_0=\theta(x=0)$ is the rescaled tilt at the surface. We next obtain the minimum $\partial_x \theta$ by first obtaining the Euler-Lagrange equation:
\begin{eqnarray}
\partial_x^2\theta= \frac{1}{2\alpha(T,\epsilon)}\frac{dg(\theta) }{d\theta}\;,
\label{E-L eqn}
\end{eqnarray}
where $g(\theta)=\alpha(T,\epsilon) \theta^2 -  \theta^4+\frac{1}{3} \theta^6$. The above equation, along with the fact that $\partial_x \theta\rightarrow0$ as $\theta\rightarrow0$, can then be integrated to yield
\begin{eqnarray}
\partial_x \theta= -\sqrt{\frac{ g(\theta)}{\alpha(T,\epsilon)}} \;,
\label{First integral}
\end{eqnarray}
where we have chosen the negative slope to ensure that $\theta(x\rightarrow\infty)=0$. Using Eq.~(\ref{First integral}) we can express the rescaled total free energy per unit area $f=f_B+f_S$ as 
\begin{eqnarray}
f= 2 \sqrt{\alpha(T,\epsilon)}\xi_x\int_0^{\theta_0} d\theta \left[\sqrt{g(\theta)}-\mu(\epsilon)\right]\;,
\label{Total free energy ito theta_0}
\end{eqnarray}
where $\mu(\epsilon)=\frac{2\sqrt{2}c s \chi\gamma}{\sqrt{K_T |b|^3}}\epsilon$ is a dimensionless measure of the strength of the coupling of tilt to the surface. Minimizing the above $f$ with respect to $\theta_0$ we find the following implicit equation for the surface tilt $\theta_0$:
\begin{eqnarray}
g(\theta_0) =\mu^2(\epsilon) \;.
\label{theta_0 condition}
\end{eqnarray}
The state map in enantiomeric excess ($\epsilon$) -- temperature ($T$) parameter space can be obtained using Eq.~(\ref{T_1st}) for $T_{1st}^*$ along with Eqs.~(\ref{Total free energy ito theta_0}) and~(\ref{theta_0 condition}), and is shown schematically in Fig.~\ref{StateMap}(a). Profiles of the surface tilt $\theta_0$ as a function of $\epsilon$, are shown in Fig.~\ref{StateMap}(b). Since the surface induced tilt generally changes very little with $T$ we do not show the surface tilt $\theta_0$ as a function of $T$.

For sufficiently large enantiomeric excess $\epsilon>\epsilon_{1st}(T)$ the system's bulk is in the Sm-$C^*$ phase. This regime lies to the right of the solid line in Fig.~\ref{StateMap}(a). For the remainder of our analysis we will focus on the region where $\epsilon<\epsilon_{1st}(T)$ corresponding to the system's bulk being in the Sm-$A^*$ phase. For sufficiently large temperatures $T>T_C$ the surface tilt $\theta_0(\epsilon)$ will increase continuously (until $\epsilon_{1st}(T)$ is reached) as $\epsilon$ is increased, as shown schematically in Fig.~\ref{StateMap}(b). For sufficiently small $T<T_C$ there will be a range of $\epsilon$ in which $\theta_0(\epsilon)$ is multivalued and the system can either be in a state of low surface tilt (LST), in which $\theta_0(\epsilon)\approx\mu(\epsilon)/\sqrt{\alpha(T,\epsilon)}$, or high surface tilt (HST). In Figs.~\ref{StateMap}(a) and (b) this coexistence region lies between the dotted lines, which represent the limits of metastability of the HST and LST states. The width of the coexistence region shrinks to zero at the critical point ($\epsilon_C$,$T_C$). While the system can exist in either the LST or HST state within the coexistence region, for smaller $\epsilon$ values the LST state will be energetically favorable while for larger $\epsilon$ values the HST will be energetically favorable. Thus, the system will jump from one state to the other at $\epsilon_J(T)$, which is defined as the $\epsilon$ value where the two states have equal free energies. This first order boundary $\epsilon_{J}(T)$, shown as a dashed line in Fig.~\ref{StateMap}(a), terminates at the critical point ($\epsilon_C$,$T_C$), where the size of the jump shrinks to zero.  The boundary  does not intersect $\epsilon_{1st}(T)$ at $\epsilon=0$ and $T=T_{1st}$, but rather at $\epsilon_L>0$ and $T_L>T_{1st}$ \cite{BECE Footnote}. Thus, only within the temperature range $T_L<T<T_C$ will one see a jump between LST and HST as $\epsilon$ is varied. For $T_{1st}<T<T_L$ the system will remain in the LST state as $\epsilon$ increases until the first order bulk Sm-$A^*$--Sm-$C^*$ boundary is reached. The location of the critical point can be found analytically using the fact that at ($\epsilon_C$,$T_C$), both the slope and curvature of the tilt profile $\theta_0(\epsilon)$ diverge, or equivalently, $\left(\frac{d \theta_0}{d \epsilon}\right)^{-1}=\left(\frac{d ^2\theta_0}{d \epsilon^2}\right)^{-1}=0$. This yields 
\begin{eqnarray}
\epsilon_C =\frac{\sqrt{K_T|b|^3}}{2\sqrt{6}c\chi s \gamma} \;
\label{epsion_C}
\end{eqnarray}
and
\begin{eqnarray}
T_C =T_{1st}+\left( \frac{T_{1st}-T_0}{3} \right)\left[1+\frac{2K_T|b|}{3cs^2\chi}\right] \;.
\label{epsion_C}
\end{eqnarray}

The surface induced tilt decays as one moves from the surface into the bulk, i.e. as $x$ increases from zero. For a given value of $\epsilon$ this decay is governed by Eq.~(\ref{E-L eqn}), or equivalently Eq.~(\ref{First integral}). The latter equation implies that $\theta(x)$  decays with increasing $x$. The length scale for this decay is $\xi_x(T)=\sqrt{K_T/a(T,\epsilon)}$ which, due to the first order nature of the bulk Sm-$A^*$--Sm-$C^*$ transition, remains finite as $\epsilon\rightarrow \epsilon_{1st}(T)$. For $T>T_c$ the function $g(\theta)$ is monotonically increasing, which implies that the curvature $\partial_x^2\theta$ is positive and the slope of $\theta$ vs $x$ becomes shallower with increasing $x$, as shown qualitatively in Fig.~\ref{StateMap}(c). For $T<T_c$, the nature of the decay can be qualitatively different. This is due to the function $g(\theta)$ being nonomonotonic, with negative slope for an intermediate range $\theta_1(T)>\theta>\theta_2(T)$. For the range $\epsilon<\epsilon_{J}(T)$, where the surface tilt $\theta_0(T)<\theta_2(T)$, the decay of the tilt will be as described above. However, for the range $\epsilon_{J}(T)<\epsilon<\epsilon_{1st}(T)$ where the surface tilt $\theta_0(T)>\theta_1(T)$ there will be a range, $x_1(T)<x<x_2(T)$ (corresponding to the range $\theta_1(T)>\theta(x)>\theta_2(T)$) over which the the curvature of the decay becomes negative. This corresponds to a kink, i.e. a slope that becomes steeper with increasing $x$. This is shown qualitatively in Fig.~\ref{StateMap}(c). It can be shown \cite{Rudquist&Saunders} that for $T\lesssim T_C$ the width of the kink, $\Delta_K(T)\equiv x_2-x_1$ grows as $\Delta_K(T)\propto\sqrt{T_C-T}$. For $T_{1st}<T<T_L$, the low surface tilt state is energetically favored for all $\epsilon<\epsilon_{1st}$ and there will be no kink.

\section{Conclusion}

We have analyzed the surface electroclinic effect (SECE) in a material that exhibits a first order bulk smectic-$A^*$ (Sm-$A^*$) -- smectic-$C^*$ (Sm-$C^*$) transition, focussing in particular on the effect of varying enantiomeric excess. We have shown that due to the first order nature of the bulk Sm-$A^*$ -- Sm-$C^*$ transition \cite{EC Footnote}, the SECE can be unusually strong and that as enantiomeric excess is varied, a jump in surface induced tilt is expected. A theoretical state map, in enantiomeric excess - temperature space, features a critical point which terminates a line of first order discontinuities in the surface induced tilt.  The decay of the surface induced tilt, as one moves from surface into bulk shows that for sufficiently high surface tilt the decay will exhibit a well defined spatial kink within which it becomes especially rapid. 

In chiral smectic liquid crystal devices, like ferroelectric liquid crystal displays, the smectic layers should ideally be homogeneously aligned perpendicular to the cell substrates. In materials with the phase sequence: isotropic--Nematic ($N^*$)--Sm-$A^*$--Sm-$C^*$  the rubbed polymer alignment layers aligns the director ${\bf \hat n}$  in the $N^*$ phase and the smectic layers are formed perpendicular to ${\bf \hat n}$ at the $N^*$--Sm-$A^*$  transition. However, when the director starts to tilt upon entry to the  Sm-$C^*$ phase there is a shrinkage in smectic layer thickness leading to an unwanted chevron structure \cite{Rieker_Chevrons}. Such chevrons and the related zig-zag defects could be avoided by using de Vries smectics which have essentially no layer shrinkage. But as de Vries materials do not exhibit a nematic phase, the initial alignment of the smectic layers is much more difficult to achieve, especially with the generally strong SECE in these materials. In order to get a homogeneous smectic alignment in device cells the rubbing directions at the two surfaces have to be carefully matched to compensate for the SECE, e.g. by cross-rubbing. Our analysis suggests that it should be possible to minimize the SECE-related alignment issues, by careful tuning of the material and cell parameters to be in the LST regime, at the isotropic to Sm-A* transition. From an applicational point of view it would also be important to investigate the switching dynamics and the long term stability of both LST and HST cells, and to see whether the LST state is stable under electric fields and temperature variations. 

KS acknowledges support from the National Science Foundation under Grant No. DMR-1005834.


\begin{thebibliography}{99}
\bibitem{415Footnote} It is possible that W415 has a Sm-$A^*$ -- Sm-$C^*$ that is tricritical or very weakly first order. Unfortunately, the nature of the bulk transition is not discussed in the article \cite{Shao&Boulder}.  However, that the transition is not first order can be inferred from the lack of an S-shaped character in the tilt versus field curves for the bulk electroclinic effect (Fig. 4 in \cite{Shao&Boulder}).
\bibitem{BECE Footnote} This contrasts with the bulk electroclinic effect in which the field $E_J(T)$, at which the bulk tilt jumps between low and high value, does go to zero at  $T=T_{1st}$.
\end{thebibliography}
\end{document}